# Evidence and Characterization of a SDW Transition in $Na_{0.75}CoO_2$ Single Crystals

B. C. Sales, R. Jin, K. A. Affholter, P. Khalifah, G. M. Veith and D. Mandrus
Condensed Matter Sciences Division, Oak Ridge National Laboratory
Oak Ridge, TN 37831-6056

**Abstract**
The magnetic, thermal and transport properties of $Na_{0.75}CoO_2$ single crystals grown by the floating zone (FZ) method are reported. Magnetic susceptibility, resistivity, magnetoresistance, and heat capacity data from these crystals indicate a bulk phase transition at $T_1 = 22$ K. These data are most consistent with the formation of an antiferromagnet spin-density-wave (SDW) at 22 K with the easy axis for magnetization nearly along the **c** axis. Weak and soft ferromagnetism is observed for applied magnetic fields less than 0.5 T, which suggests a slight canting of the SDW magnetization with respect to the **c** axis. The jump in the heat capacity at the SDW transition is 0.45 J/K-mole-Co or about 50% of the value expected from mean-field weak-coupling theory. The reduced jump and the decrease in the resistivity below $T_1$ are consistent with a gap for only part of the Fermi surface. The magnetoresistance is small at the SDW transition but increases in both directions reaching a value of 100% at 2 K for applied fields of 8 Tesla. The magnetoresistance data imply that the mobility of the remaining carriers is large and increases below $T_1$. The observation of a SDW transition in this material is found to be sensitive to the preparation conditions and the degree of order in the Na layers. No SDW transition is observed in our polycrystalline powder with the same nominal composition ($Na_{0.75}CoO_2$) and lattice constants. Differential scanning calorimetry data, however, show distinct differences between the powder and crystal, suggesting a higher degree of order in the Na layers within the crystal. The crystal exhibits a sharp phase transition at $T_2 \approx 340$ K while for the powder this transition is smeared over the temperature range from 250 – 310 K.

## I. Introduction

The layered transition metal oxides, $Na_xCoO_2$, have been of considerable interest to the scientific community for the past few years. This family of materials has been known for some time [1] but the report by Terasaki et al. [2] concerning possible thermoelectric applications has rekindled interest in these compounds [3-8]. It is very unusual for an oxide to have a relatively high thermopower yet a low and metallic resistivity [9]. This result alone suggests interesting electronic correlations. The subsequent discovery of superconductivity at 5 K by Takada et al. [10] in a hydrated version of this material ($Na_{0.35}CoO_2 \cdot 1.3$ $H_2O$) caused even greater excitement [11-15]. Although the superconducting transition temperature is too low for applications, the structural and chemical similarities between $Na_xCoO_2$ compounds and cuprate superconductors provide hope of additional insight into the physics of superconductivity in layered transition metal oxides. In $Na_xCoO_2$, the Co ions in each plane form a triangular lattice rather that the square lattice found in the cuprates. A S=1/2 triangular lattice is of great interest to theorists because of the inherent geometrical frustration of antiferromagnetic order and since Anderson's original RVB model for superconductivity [16] was calculated for a triangular lattice.

The basic structure of $Na_xCoO_2$ (0.3 < x < 0.9) consists of planes of edge-sharing $CoO_6$ octahedra alternating with layers of Na. In this article we focus on the most Na rich composition with the gamma phase: $Na_{0.75}CoO_2$. The gamma phase is hexagonal ($P6_3/mmc$) with **a** = 2.83 Å and **c** ≈ 10.85 Å. There are four crystallographically distinct positions within the unit cell: Co at (0,0,0.5), O at (0.333, 0.667, 0.0913) and two Na positions Na1 (0,0,0.25) and Na2 (0.667, 0.333, 0.25). Both Na positions are only partially occupied, and cannot be occupied randomly since that would result in some of the Na at site 1 being too close to Na at site 2 [6]. One neutron structure refinement [17] estimates the average relative occupancy of Na1 ≈ 0.25 and Na2 ≈ 0.5. A Na at site 1 has a Co ion directly above and below it at a distance of 2.71 Å as compared to a Na2-Co minimum distance of 3.17 Å. In an ionic picture, the Co ions are mixed-valent with a formal oxidation state of 4-x. A simple electrostatic model would suggest that the Co near an occupied Na1 site would have a greater tendency toward $Co^{+3}$. Within this simple picture there are clearly strong tendencies toward some type of partial charge and Na ordering. Recent NMR data from $Na_{0.75}CoO_2$ powder, and resistivity and electron diffraction data from $Na_{0.5}CoO_2$ crystals provide strong evidence of charge ordering near and above room temperature [18,19].

Magnetic data [20] and electronic structure calculations [6] suggests a low-spin configuration of S=0 for $Co^{+3}$ and S = 1/2 for $Co^{+4}$ within a $t_{2g}$ ground state. Band structure calculations indicate that the six bands comprising the $t_{2g}$ manifold are split further in the rhombohedral crystal field into two $a_{1g}$ and four $e_g'$ symmetry bands. At the Fermi energy most of the bands have $a_{1g}$ character with some $e_g^1$ admixture. The calculated Fermi surface has large cylindrical hole Fermi surfaces around the zone center (Γ) of dominant $a_{1g}$ character and small holelike sections with some $e_g^1$ character centered about 2/3 of the way out along the Γ−K and A-H directions. The large cylindrical Fermi surface is flattened along the Γ−K directions resulting in a cylinder with six facets (allen wrench). Large portions of this Fermi surface are susceptible to nesting [6]. In addition, because of two dimensionality, the smaller Fermi regions may also have significant nesting between themselves. Carriers with more $e_g^1$ character tend to have lower effective masses and higher mobilities than carriers with $a_{1g}$ character.

The present article focuses on the magnetic, transport, and thermodynamic properties of single crystals of $Na_{0.75}CoO_2$ grown by a floating zone method. Two distinct phase transitions are observed in these crystals, one at $T_1$ = 22 K and the other at $T_2$ ≈ 340 K. The transition at 22 K was first reported by Motohashi et al. [21] on powder samples of $Na_{0.75}CoO_2$ and later investigated by Sugiyama et al. [22] using muon spin rotation (μsr). Motohashi et al.[21] suggested that the transition at 22 K was likely a SDW, a conclusion consistent with the μsr results. A transition at $T_2$ ≈ 340 K has not been previously reported, although it is probably related to the smeared transition between 250-310 K studied by Gavilano et al. [19] on powder and the transition at ≈430 K observed by Wang et al. [23] in flux grown crystals.

II. Synthesis and Experimental Methods

High purity powders of $Na_2CO_3$ (99.997% from Alfa-AESAR) and $Co_3O_4$ (99.999% from Alfa-AESAR) were carefully weighed to give the desired stochiometry of $Na_{0.75}CoO_2$. The powders were first mixed by hand to yield a homogenous gray color and then ball milled for 2 h. The ball milled powder was loaded into a high purity (99.9%) alumina tray, inserted into a furnace that had been preheated to 750 °C and left for 20 h. This rapid-heat technique was found to minimize the loss of Na due to volatilization [24]. The prereacted powder (total weight of about 30 g) was pressed in to three 1"diameter pellets and placed on a thin layer of the same powder in an alumina tray. The pellets were heated to 830 °C for 16 h in pure $O_2$ that was slowly flowed over the tray (≈ 10 cc/min), and then cooled to room temperature over a period of 6-8 h. The final powder was obtained after ball milling all three 1" pellets for an additional 1-2 h. Measurements of the weights before and after heat treatments indicated that the final stochiometry was very close to $Na_{0.75}CoO_2$. ICP (Inductively-coupled-plasma) analysis of the powder gave a Na/Co ratio of 0.74±0.02. Idiometric titration measurements gave a formal Co valence of 3.25 ±0.01 in good agreement with the ICP and weight loss data. Powder x-ray diffraction showed only the desired hexagonal gamma phase *P6₃/mmc* with **a** = 2.83 Å and **c** = 10.85 Å.

Single crystals with a composition close to $Na_{0.75}CoO_2$ were prepared from the $Na_{0.75}CoO_2$ powder using a floating zone method. The powder was hydrostatically compressed in rubber bladders into 6 mm diameter by 100 mm long rods. The fragile rods were carefully removed from the bladder, placed in an alumina tray on loose powder with the same composition, and heated at 830 °C for 16 h in flowing oxygen. The polycrystalline rods were used in an NEC SCM15-HD Arc Image Furnace to prepare single crystals of $Na_{0.75}CoO_2$ via the floating zone method. Laue x-ray photographs were used to cut or cleave oriented single crystal plates with typical dimensions of 5 x 5 x 1 mm$^3$ from the as-grown boule. The crystal plates had lattice constants of **a** = 2.83 Å and **c** = 10.86 Å, very similar to the values found for the starting powder. Both the powder and the crystals would react with moisture and $CO_2$ in the air if left exposed to ambient air for several days. A small amount of white powder would sometimes form on the surface of the crystal. This powder was identified as $Na_2CO_3 \cdot yH_2O$.

For comparison purposes, crystals of $Na_xCoO_2$ were grown from a NaCl flux in an alumina crucible as described in [5].These crystals typically contain small amounts of Al and perhaps other impurities. We found that the physical properties of the flux-grown crystals are significantly different at both low and high temperatures from crystals prepared with the floating zone method. The typical lattice constants from the flux grown crystals are **a** = 2.83 Å and **c** = 10.92 Å. The value for the **c** lattice constant indicates an average composition of $Na_{0.72}CoO_2$ [18].

We also tried to prepare powder with Na contents of x=0.6 and x = 0.67 using the same protocol developed for preparing the x = 0.75 powder. Both the x=0.6 and 0.67 powders showed significant amounts of $Co_3O_4$ either in the x-ray pattern or in the magnetic susceptibility data. If the powders are heated at still higher temperatures (≈ 900- 950 °C) the $Co_3O_4$ impurity converted or partially converted to a CoO impurity phase. We conclude that for synthesis methods in the 700-950 °C temperature range, the gamma phase only forms for a small range of Na concentrations x = 0.75 ± 0.03. Gamma phase samples of $Na_xCoO_2$ with lower concentrations of Na, however, can be prepared using non-equilibrium room temperature chemical or electrochemical extraction methods [10, 25].

A powdered sample of $Na_{0.37}CoO_2$ was prepared by stirring 8 g of the $Na_{0.75}CoO_2$ powder in a 6.6 M bromine/ acetonitrile solution (15 cc 99.99% Br and 75 cc 99.999% $CH_3CN$) for 5 d.  The powder, which is very hydroscopic, was thoroughly washed with pure $CH_3CN$ and quickly transferred to a vacuum chamber where the small amount of remaining $CH_3CN$ was pumped away resulting in a dry and fluffy powder. ICP and iodiometric titration analysis of the powder yielded a Na content of x = 0.37 ± 0.02. Powder x-ray diffraction yielded **a** = 2.83 Å and **c** = 11.20 Å. This value for **c** is consistent with a Na content of x = 0.37 [18].

X-ray diffraction measurements were made with a SCINTAG powder diffractometer using Cu Kα radiation. Patterns were calibrated with a $LaB_6$ standard through the JADE software package. The procedures used for the idiometric titration measurements have been described previously [26]. Scanning calorimetry measurements were made from 120-500 K using a Perkin-Elmer DSC-4. Magnetic susceptibility and magnetization data were collected using a commercial SQUID magnetometer from Quantum Design. Resistivity, heat capacity, thermal conductivity and Seebeck measurements were made using the PPMS (Physical Property Measurement System) from Quantum Design. To achieve a low contact resistance, gold pads (≈1000 Å thick) were sputtered onto the crystals. Electrical leads were then attached to the gold pads using silver epoxy( H20E from EPOTEK).

### III. Magnetic Susceptibility and Scanning Calorimetry

Magnetic susceptibility data are shown in Fig 1 for the starting powder ($Na_{0.75}CoO_2$), floating zone (FZ) crystals ($Na_{0.75}CoO_2$) and flux grown crystals ($Na_{0.72}CoO_2$). For comparison purposes, the data for the crystals is the polycrystalline average for the two directions, $\chi_{av} = (2\chi_a + \chi_c)/3$.  These data illustrate how the properties of the material can vary with small changes in either Na stochiometry or disorder within the Na layer. Only the susceptibility data from crystals prepared by the floating zone method show a clean magnetic transition at 22 K. For temperatures above 50 K, all of the susceptibility data in Fig 1 can be parameterized by $\chi = \chi_o + C/(T+\Theta)$. Least squares fits of the data gives effective moments of 1.12-1.36 $\mu_B$ per Co, and values of $\chi_o$ between 1.36 x $10^{-4}$ and 1.71 x $10^{-4}$ $cm^3$/mole. The constant $\chi_o$ accounts for Pauli paramagnetism, core diamagnetism and Van Vleck contributions from the Co ions. The Van Vleck contributions aree known to be large in many Co compounds [27]. The Curie-Weiss temperatures, $\Theta$, are 120 K,

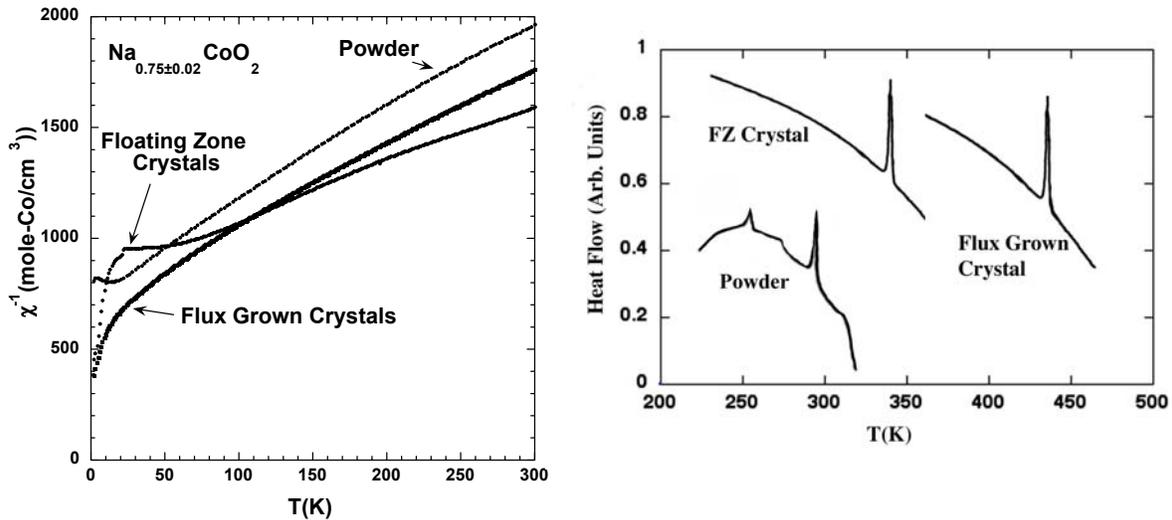

Fig.1(Left) Inverse magnetic susceptibility vs. temperature for three samples of $Na_{0.75 \pm 0.02}CoO_2$ prepared by different methods. These data were taken in an applied field of 0.1 Tesla.(Right) DSC scans from same three samples.

122 K, and 206 K for the powder, flux grown crystal, and FZ crystal, respectively. The larger value of the Curie-Weiss temperature for crystals grown by the floating zone method may indicate either stronger antiferromagnetic interactions or increased mixing between quasi-localized Co d orbitals and conduction electrons.

Since the powder and FZ crystals have virtually identical chemical compositions, the likely origin of the difference in the magnetic properties is the degree or type or order in the Na layers between the $CoO_2$ planes. An initial analysis of x-ray powder diffraction data showed no obvious structural differences between the two materials. Scanning calorimetry data, however, showed clear differences. The powder exhibited two to four distinct endothermic peaks when heating through the temperature range from 250-310 K. The heat associated with each peak ranged from $\approx$ 100-400 J/mole $Na_{0.75}CoO_2$. In the gamma structure there are two crystallographically distinct Na sites with each site only partially occupied [17]. It is likely that the endothermic peaks correspond to a rearrangement of the Na within each layer and possibly the onset of charge ordering in the Co layers. DSC measurements on the FZ crystals showed only one sharp endothermic peak at 343 K with a total heat absorption of about 1500 J/mole $Na_{0.75}CoO_2$. Measurements on several different crystals indicate a strong correlation between a sharp DSC peak at 343 K and a distinct magnetic transition at 22 K. DSC measurements were also performed on several flux grown crystals ($Na_{0.72}CoO_2$). These crystals exhibited a sharp endothermic peak at 430 K with a heat content of about 2400 J/mole. However, there was no distinct magnetic transition for these crystals, possibly because small amounts of impurities incorporated during the growth process smear the transition. DSC measurements on the $Na_{0.37}CoO_2$ powder showed no thermal anomaly in the temperature range from 120-500 K.

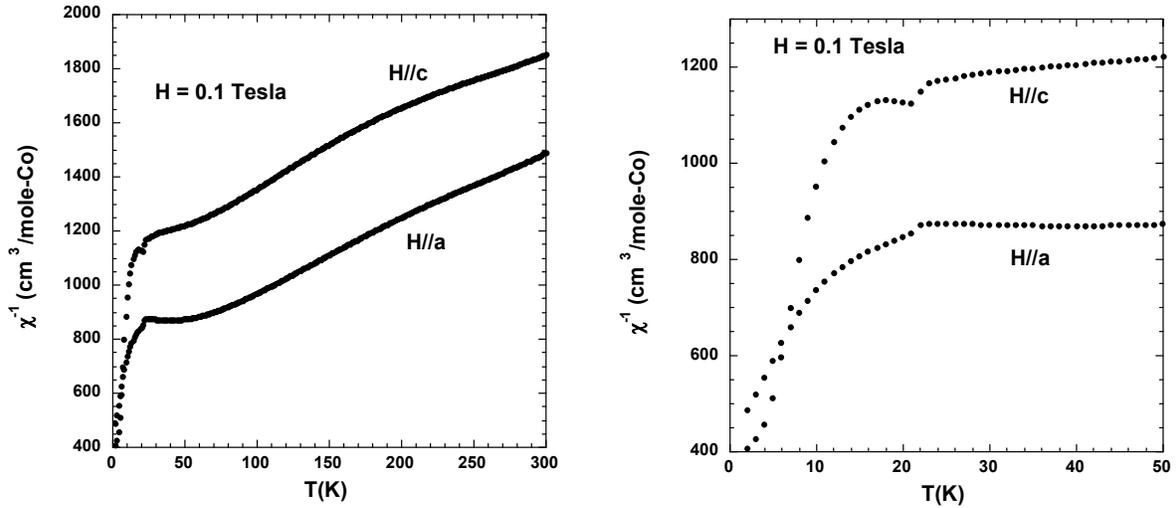

Fig 2 (Left) Inverse magnetic susceptibility versus temperature for a $Na_{0.75}CoO_2$ crystal grown via the floating zone method. (Right) Low temperature plot of same data. A magnetic transition at 22 K is evident for H // **c** and H // **a**.

The remainder of this article will focus on the properties of $Na_{0.75}CoO_2$ crystals prepared using the floating zone method and the nature of the magnetic transition at 22 K and the Na or charge ordering transition at 343 K. The magnetic susceptibility data for magnetic fields applied parallel to **a** and **c** are displayed in Fig 2. There is a clear and abrupt increase in the susceptibility in both directions at $T_1$ = 22 K for in a field of 0.1 Tesla. This transition was first reported by Motahashi et al. [21] in $Na_{0.75}CoO_2$ powder. Above 50 K the susceptibility data in Fig 2 can be parameterized by $\chi = \chi_o + C/(T+\Theta)$ with $\chi_o$ = 9.49 x $10^{-5}$ $cm^3$/mole Co, $p_{eff}$ = 1.56 $\mu_B$, and $\Theta$ = 220 K for H//**a** and $\chi_o$ = 2.76 x $10^{-4}$, $p_{eff}$ = 1 $\mu_B$, and $\Theta$ = 167 K for H//**c**. The anisotropy in the susceptibility data is significant ranging from 25% at 300 K to 45% at 50 K. Magnetization data at 1.9 K are shown in Fig. 3 for H // **a** and H // **c**. In both directions there is evidence of weak ferromagnetism and hysteresis for applied magnetic fields of less than 0.5 Tesla. The remanent magnetization is very small and corresponds to about 0.0001 $\mu_B$ per Co for H // **c** and less for H // **a**. The coercive field is about 100 Oe. For applied fields larger than 1 or 2 T the magnetization is linear in field for H// **a** and nearly linear for H // **c**. The magnetization data for H//**c** begins to curve upwards for H > 5 T, which suggests a possible metamagnetic transition at fields greater than 7T (the limit of our SQUID). The extremely small magnetization associated with the ferromagnetic behavior at low fields, coupled with the linear and superlinear behavior of the magnetization at higher fields
suggests that ferromagnetism is not the dominant magnetic response of the system. To test this hypothesis the susceptibility of $Na_{0.75}CoO_2$ was measured in an applied field of 5 Tesla (Fig 4). The susceptibility data resembles that of an antiferromagnet [28] with the magnetic spins aligned along the **c** axis of $Na_{0.75}CoO_2$. The increase in the susceptibility below 22 K for H // **a**, suggests that the magnetization density for each sublattice may be tilted slightly with respect to the **c** axis. In antiferromagnetically ordered insulating compounds, such as $K_2V_3O_8$ [29] this tilting results in weak ferromagnetic behavior, similar to that shown in Fig 3. In many insulating weak ferromagnets (including $K_2V_3O_8$) the tilting is due to the Dzyaloshinskii-Moriya interaction. For comparison, the magnetization data at 1.9 K for the $Na_{0.75}CoO_2$ powder (not shown) is perfectly linearly for fields from 0 to 7 T.

Magnetic susceptibility data were also taken for temperatures between 300 and 400 K to see if the transition at 340 K was magnetic. There was a very small increase (jump of about 1%) in the magnetic susceptibility on cooling below 340 for H//**c** (Fig. 5) and an even smaller effect for H // **a**.

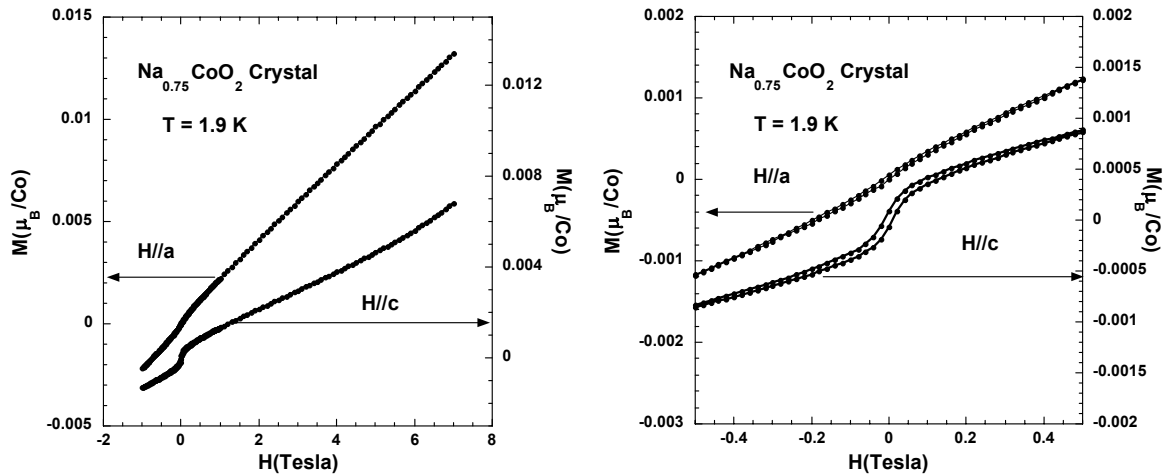

Fig. 3 (Left) Magnetization versus field at T = 1.9 K for a $Na_{0.75}CoO_2$ crystal grown via the floating zone method. Very weak ferromagnetic behavior is evident for H < 0.5 T. At higher fields, the magnetization data are more consistent with a type of antiferromagnetic ground state. (Right) Low-field plot of data shown on left.

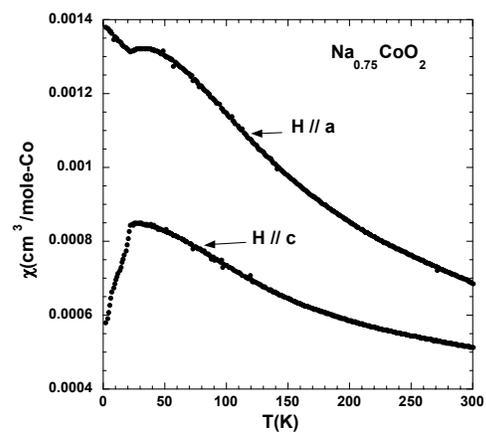

Fig. 4 Susceptibility versus temperature with an applied field of 5 Tesla.

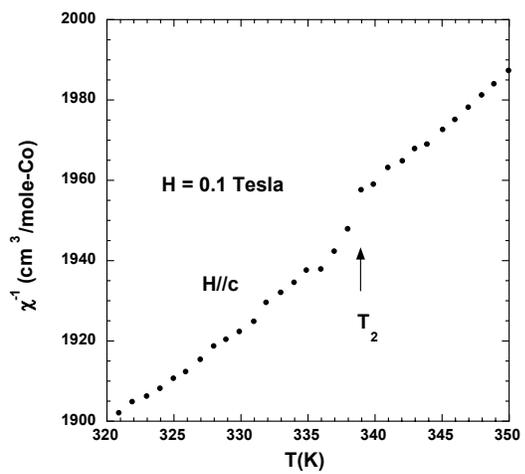

Fig. 5 Inverse susceptibility of $Na_{0.75}CoO_2$ versus temperature in the vicinity of $T_2$. There is about a 1% change in $\chi$ near $T_2$.

## IV. Resistivity and Magnetoresistance

The temperature dependence of the resistivity data are shown in Fig 6a. Both the in-plane, $\rho_a$, and out-of-plane, $\rho_c$, resistivities are metallic over the entire temperature range (2-400 K) with $\rho_c$ about 600 times larger than $\rho_a$. The Na and possible partial charge ordering transition at 340 K ($T_2$) and the magnetic transition at 22 K ($T_1$) are clearly evident in the resistivity data. Cooling the crystal through each transition results in an abrupt decrease in the resistivity. By 2 K the resistivity in both directions is 10 times smaller than the value extrapolated from data above 22 K. This strongly suggests that the magnetic phase transition that occurs at $T_1$ is a bulk effect and is not due a small amount of a magnetic impurity phase. The magnetoresistance is small near $T_1$ but rapidly increases as the temperature is reduced further (Fig. 6b). This behavior is inconsistent with an onset of weak ferromagnetism below $T_1$. For ferromagnets the magnetoresistance is maximal near the ordering temperature in sharp contrast to the behavior of $Na_{0.75}CoO_2$. The field dependence of $\rho_c$ at 2 and 5 K for H // **c** is shown in Fig 7a. At 2 K a field of 8T more than doubles the resistivity. In spite of the large magnetoresistance at low temperatures, a field of 8T has no detectable effect on the transition temperature of 22 K. In "normal" metals the magnitude of the magnetoresistance increases as the mobility of the carriers increases (or as the resistivity decreases). This is formalized as Kohler's rule [30], which states that the size of the magnetoresistance is a function the applied field divided by the resistivity in zero field {i.e. $(\rho(H,T)-\rho(0,T))/\rho(0,T) = F (H/\rho(0,T))$}. Motivated by this relation, we show in Fig. 7b that the magnetoresistance below $T_1$ scales as the conductivity $(1/\rho(0,T))$. This suggests that the large magnetoresistance below $T_1$ is due to an increase in carrier mobility.

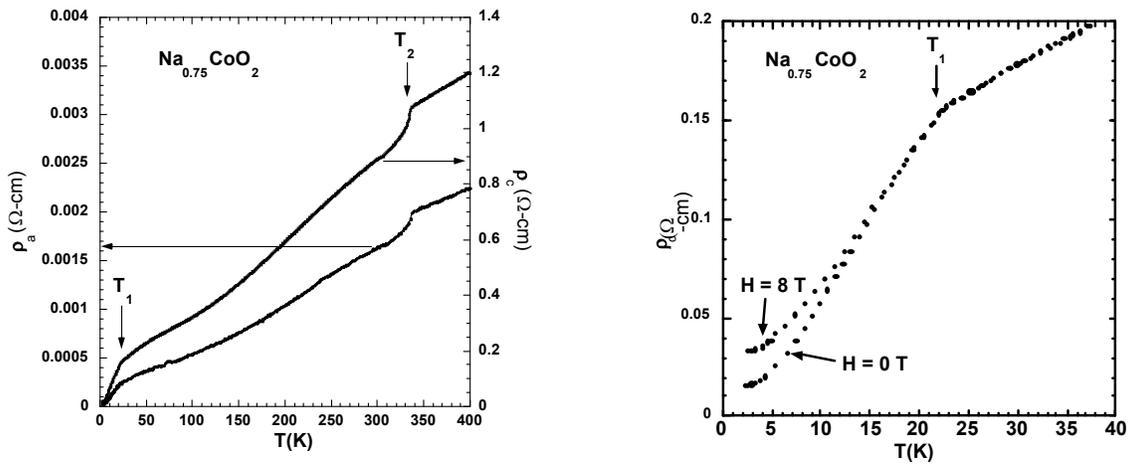

Fig. 6. (Left) Resistivity of a $Na_{0.75}CoO_2$ crystal for current in the plane ($\rho_a$) and along the c axis ($\rho_c$). The magnetic transition at $T_1 = 22$ K and the ordering transition at $T_2 = 340$ K are evident in the resistivity data. (Right) Low temperature resistivity data ($\rho_c$) versus temperature in 0 and an 8 Tesla magnetic field (H// **c**). Similar (but noisier) data were also found for $\rho_a$ with H // **c**.

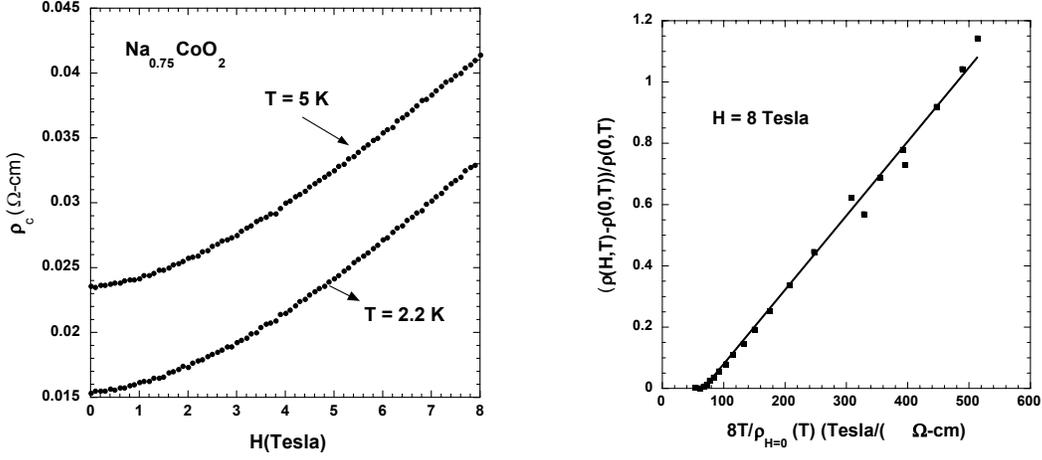

Fig. 7 (Left) Resistivity ($\rho_c$) versus magnetic field (H // **c**) at T= 2.2 K and T = 5 K. (Right) Magnitude of magnetoresistance at 8 Tesla versus conductivity $(1/\rho(T)_{H=0})$ at zero field. This plot was motivated by "Kohler's Rule" [30] and suggests that the large magnetoresistance at low temperatures is due to an increase in the mobility of the carriers as the temperature is lowered below $T_1$.

**V. Heat Capacity**

Heat capacity data versus temperature for a $Na_{0.75}CoO_2$ crystal and for the $Na_{0.75}CoO_2$ starting powder are shown in Fig. 8. As discussed in Section II, we believe that the only significant difference between the powder and the crystal is the degree of order in within the Na layers. The heat capacity data from the crystal exhibits a distinct transition at 22 K that is very similar in shape to the heat capacity anomaly from a BCS superconductor. The heat capacity data from the powder exhibits no phase transition in the temperature range from 5 to 30 K and can be accurately described as a sum of an electronic and lattice contribution ($C_{powder} = \gamma T + \beta T^3$) with $\gamma = 0.027$ J/$K^2$-mole-Co and $\beta = 4.3384 \times 10^{-5}$ J/$K^4$-mole-Co. This value for $\beta$ implies a Debye temperature, $\Theta_D$, of 545 K. The values for $\gamma$ and $\Theta_D$ are similar to the values estimated by Motohashi et al.[21]. The data from the powder was scaled by a few percent to exactly match the heat capacity of the crystal at 30 K. The weight of each sample is typically only accurate to within 2-5%. The entropy associated with the magnetic transition is estimated to be only 0.08 J/mole-Co-K. If this transition corresponded to the long range magnetic ordering of localized $Co^{+4}$ spins, we would expect the entropy associated with the transition to be of order $0.25R\ln(2) \approx 1.44$ J/mole-Co-K, which is 20 times larger than found experimentally. The magnitude and shape of the heat capacity feature near 22 K (Fig 8), however, is consistent with the formation of a gap over part of the Fermi surface as a result of the development of a spin-density-wave (SDW). In the simplest models of a SDW, the transition is considered to involve only the electrons near the Fermi energy. The lattice contribution to the heat capacity from the crystal was estimated using the data from the powder, which doesn't show a transition. The electronic portion of the heat capacity divided by T ($C_{el}/T$) for the crystal and the powder are shown in Fig 9. For a SDW, the net entropy change below 22 K associated with the transition must coincide with the entropy reduction if the electronic density of states maintains its value just above the transition ($\gamma = 0.027$ J/K$^2$- mole Co) all the way to T= 0 [31]. This constraint means that the area of the peak with $C_{el}/T > \gamma$ has to equal the area where $C_{el}/T < \gamma$. Based on this constraint we conclude that the downturn at about 6 K in the $C_{el}/T$ data from the crystal is not associated with the SDW transition but may be extrinsic and sample dependent. The expected behavior of $C_{el}/T$ below 6 K due to the SDW transition is sketched in Fig 9. This extrapolation is consistent with the entropy constraint. The jump in the electronic portion of the heat capacity at the SDW transition is $\Delta C = 0.45$ J/mole-Co-K and hence $\Delta C/\gamma T_{SDW} = 0.75$. In the mean-field weak-coupling-limit theory predicts that this ratio should be the same as that for a BCS superconductor, $\Delta C/\gamma T_{SDW} = 1.43$ [32]. Since the resistivity data clearly indicate that the entire Fermi surface is not gapped, the ratio $0.75/1.43 = 0.52$ can be used as a crude estimate of the fraction of the Fermi surface affected by the SDW transition. The heat capacity of the crystal was also measured in a field of 8 T with H // **c** (not shown). There was no detectable effect of a magnetic field on the data shown in Fig. 8.

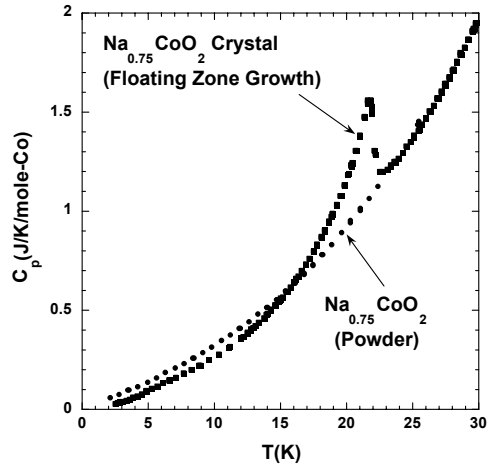

Fig.8. Heat capacity versus temperature for a crystal prepared via the floating zone method and powder with essentially the same composition. As discussed in section II, the only significant difference between the two samples appears to be the degree of order in the Na layers. This difference is clearly reflected in DSC data taken between 100 K and 500 K.

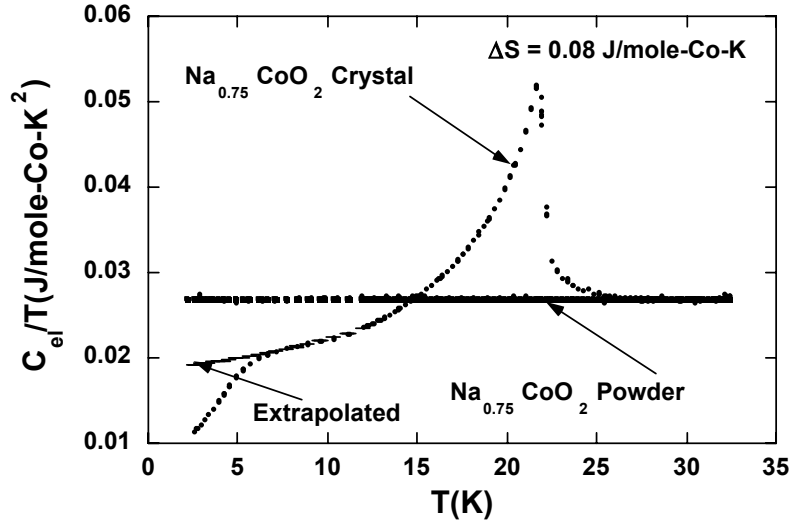

Fig, 9. Electronic portion of the heat capacity divided by temperature versus temperature for the crystal and the powder. The shape of the peak in the heat capacity data and the magnitude of the entropy associated with the magnetic transition at 22 K strongly imply the formation of a SDW. The downturn in $C_{el}/T$ at about 6 K for the data from the crystal is most likely extrinsic. The extrapolated behavior shown for $C_{el}/T$ is consistent with the conservation of entropy.

## VI. Thermal Conductivity and Thermopower

The thermal conductivity, $\kappa_a$, and thermopower, S, data from a $Na_{0.75}CoO_2$ crystal are shown in Figs. 10 and 11 with the temperature gradient and heat flow along the **a** axis. The temperature dependence of the thermal conductivity data (Fig 10a) is typical of an insulating crystal, which implies that most of the heat in $Na_{0.75}CoO_2$ crystals is carried by phonons. The lack of a clear anomaly at 22 K implies relatively weak coupling between acoustic phonons that carry most of the heat and the electrons involved in the SDW transition. As has been pointed out by several authors [2,8] the thermopower of $Na_{0.75}CoO_2$ near room temperature is unusually high (S ≈ 115 μV/K) for a metal, which has led to interest is this and related layered $CoO_2$ compounds for thermoelectric applications. The rearrangement of the electronic structure that occurs near the SDW transition at 22 K has a clear impact on the thermopower below 22 K. A qualitative discussion of of S below 22 K, however, requires detailed knowledge of the new electronic structure.

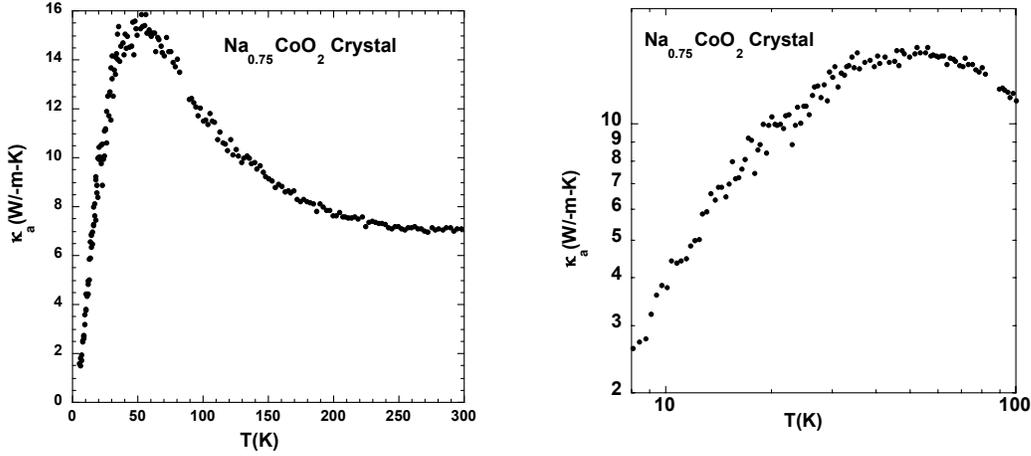

Fig 10. (Left) In-plane thermal conductivity, $\kappa_a$, versus temperature for a $Na_{0.75}CoO_2$ crystal. (Right) Log of $\kappa_a$ versus log of temperature. Notice that there is no clear sign of the SDW transition at 22 K in the thermal conductivity data. This result implies weak coupling between the conduction electrons involved in the SDW and the acoustic phonons that carry most of the heat.

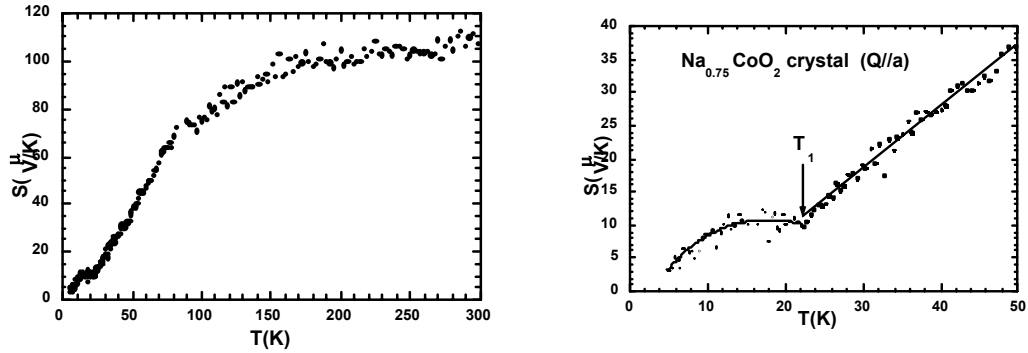

Fig. 11. Thermopower, S, versus temperature. (Left) The value of S at high temperatures is unusually high for a metal, which has attracted interest in this material for thermoelectric applications.(REFS) (Right) The SDW transition at $T_1 = 22$ K has a clear effect on the temperature dependence of the thermopower, consistent with a modification of the structure in the density of states near the Fermi energy.

## VII. Summary and Conclusions

The properties of $Na_{0.75}CoO_2$ are sensitive to the details of sample preparation. Polycrystalline powder and single crystals of $Na_{0.75}CoO_2$ with the same nominal compositions and lattice constants exhibit qualitatively different properties. The crystals exhibit a SDW transition at 22 K, and a sharp first order transition at about 340 K with a latent heat of 1500 J/mole. The powder shows no magnetic transition above 2K and the first order transition at higher temperatures is smeared over the temperature range from 250-310 K as a series of first order transitions with latent heats ranging from 100-400 J/mole. The variation in properties is likely related to the distribution of Na in each Na layer. The two Na sites in the structure (Na1 at 0,0,0.25) and (Na2 at 0.667, 0.333, 0.25) are partially occupied with the Na1 site about 1/4 filled and the Na2 site about 1/2 filled [17]. The two sites cannot be occupied randomly since that would result in some of the Na at site 1 being too close to the Na at site 2. In addition, a Na in the Na1 site is substantially closer to a Co ion than Na in site 2, which means that the Co directly above or below an occupied Na1 site will have a greater tendency toward $Co^{+3}$. Within this simple picture there are clearly strong tendencies toward some type of charge and Na ordering. As to the low temperature ground state, LSDA calculations predict a weak ferromagnetic instability, but Singh [6] notes that close in energy there are " a large space of spin and charge orderings that are nearly degenerate with each other".

The crystals exhibit a phase transition at $T_1$ = 22 K. Heat capacity, magnetization, resistivity, and magnetoresistance data are most consistent with a SDW phase transition at 22 K with a gap that develops over only a portion of the Fermi surface. The heat capacity data exhibit a feature at 22 K that is very similar in shape to the heat capacity peak from a BCS superconductor. The entropy associated with the transition is small, only 0.08 J/mole-Co. The jump in the heat capacity at $T_1$ is 0.45 J/K-mole Co or about 50% of the value expected from mean-field theory. Low and high field magnetization data indicate that the easy axis for the spin density magnetization is nearly along **c**. Upward curvature in the high-field magnetization data at 2 K implies that a spin-flop transition occurs for H//**c** for fields slightly higher than 7 T (the limit of our instrument). Weak and soft ferromagnetism for fields less than 0.5 T, suggests a slight canting of the SDW magnetization with respect to the **c** axis. The temperature of the SDW ($T_1$ = 22K) transition is not changed in an 8 T magnetic field, but there is a huge magnetoresistance at low temperatures; the resistivity at 2 K doubles in a field of 8 T. The resistivity in both directions ($\rho_a$, $\rho_c$) decreases more rapidly below 22 K than expected from an extrapolation of the resistivity data above 22 K. As discussed in section I, the calculated Fermi surface has two hole-like parts. The larger portion of the Fermi surface consists of a faceted cylinder (allen wrench) of mainly $a_g$ character. The smaller holelike sections have both $a_g$ and $e_g^1$ character. Nesting and the formation of a SDW are most likely to occur for the faceted cylinder with a nesting vector of 4/5 K [6]. Holes from the faceted portion of the Fermi surface have high effective masses and may act as strong scattering centers for the lighter carriers from the smaller holelike sections. This scenario would explain why the resistivity decreases below 22 K even though the total number of carriers is lower.

The crystals also exhibit a first-order phase transition at $T_2 \approx 340$ K with an associated latent heat of 1500 J/mole $Na_{0.75}CoO_2$. The resistivity exhibits a rapid decrease as the crystal is cooled through the transition suggesting reduced scattering due to a higher degree of order. There is very little change ( less than 1%) in the magnetic susceptibility near $T_2$ and no obvious change in the crystal structure. The transition probably involves ordering within the Na layers with perhaps some partial charge ordering within the Co layers. A detailed Na NMR experiment on a powdered sample with the same nominal composition ($Na_{0.75}CoO_2$) interpreted the smeared transition from 250-310 K as evidence for charge ordering [19]. In addition, recent electron diffraction and resistivity data on a $Na_{0.5}CoO_2$ crystal (prepared using Br to remove some Na from a $Na_{0.75}CoO_2$ crystal) showed clear evidence for charge ordering at room temperature [18] . These data clearly indicate the need for a more complete investigation of the transition at 340 K.

**VIII Acknowledgements**

It is a pleasure to acknowledge stimulating discussions with David Singh. Oak Ridge National Laboratory is managed by UT-Battelle, LLC, for the U. S. Department of Energy under Contract No. DE-AC05-00OR22725.